# Surface Plasmon Electrochemistry

Zohreh Hirbodvash,[1,2] and Pierre Berini[1,2,3]*
[1]Dept. of Physics, University of Ottawa, 150 Louis Pasteur, Ottawa, Ontario, K1N 6N5, Canada
[2]NEXQT Institute, University of Ottawa, 25 Templeton St., Ottawa, Ontario, K1N 6N5, Canada
[3]School of Electrical Engineering and Computer Science, University of Ottawa, 25 Templeton St., Ottawa, Ontario, K1N 6N5, Canada
*Corresponding author email: berini@eecs.uottawa.ca

**Abstract**

Surface plasmon polaritons (SPPs) are optical waves that propagate along a metal surface. They exhibit properties such as sub-wavelength localization and field enhancement which make them attractive for surface sensing, as commonly encountered in surface plasmon biosensors - the most widespread of all optical biosensors. Electrochemistry also occurs on metal surfaces, and electrochemical approaches are widely used to implement biosensors - electrochemical biosensors are the most prevalent biosensors in use. Given that metal surfaces are inherent to both techniques, it is natural to combine them into a single platform. The motivation may be (i) to realise a multimodal biosensor (electrochemical, optical), (ii) to use SPPs to probe electrochemical activity or the electrochemical double layer, thereby revealing additional or complementary information on the redox reactions occurring thereon, or (iii) to use SPPs to affect (pump) electrochemical reactions with non-equilibrium energetic (hot) electrons and holes created in working electrodes by SPP absorption, potentially leading to novel redox reaction pathways (plasmonic electrocatalysis). We introduce in a tutorial-like fashion basic concepts related to SPPs on planar structures and to electrochemistry, then we review non-exhaustively but representatively literature on the integration of these techniques.

## 1. Introduction

Analyzing biomolecules is required in many fields ranging from food safety inspection to medical diagnostics [1,2,3]. Many labelled detection methods have been used to detect biomolecules. These labeled detection techniques include enzyme-linked immunosorbent assay (ELISA) [4,5], colorimetric and fluorescence detection [6,7,8], polymerase chain reaction (PCR) [9,10], radioactive isotopes (radio-immunoassays) [11], vibrational spectroscopy (*e.g.*, infrared and Raman spectroscopy) [12] and some other techniques [13,14]. Labelled techniques generally create liabilities due to the use of labels. An example of a liability created by labelling is interference of the fluorophore with binding kinetics, and changes in fluorescence dynamics depending on specific dye–protein interactions after being exposed to analytes [15]. Such liabilities can be eliminated by using label-free methods, as have been employed to detect biomolecules using mass spectrometry (MS) [16], quartz crystal microbalance (QCM) [17], surface plasmon resonance (SPR) [18], localized surface plasmon resonance (LSPR) [19], long-range surface plasmon polaritons (LRSPPs) [20], and anomalous reflection of gold [21].

A biosensor is a transduction device that is used to measure biological or biochemical reactions by generating quantitative signals (electrical, thermal, optical) proportional to the concentration of an analyte [2,22,23]. A biosensor has a recognition element (*e.g.*, enzymes, nucleic acids, cells, and micro-organisms or antibodies) to selectively capture the analyte of interest in a sample. Depending on the underlying transducer technology, one can identify several types of biosensors: electrochemical [24,25], piezoelectric [26,27], thermal [28], and optical [29].

The most common type of biosensor is electrochemical [30-33]. Electrodes have an essential role as solid support for the immobilization of biomolecules and electron transfer to/from the redox species. As the result of certain electroactive species undergoing a redox reaction in the system, a voltage or a current is generated. One of the critical aspects of an electrochemical biosensor is the use of enzymes as predominant recognition elements due to their specific binding capabilities and biocatalytic activity [31,32]. In this sensing method, different electrochemical detection techniques such as amperometric, potentiometric, and conductometric techniques are used [33].

A piezoelectric sensor uses the piezoelectric effect, and the QCM is the most well-known type of piezoelectric sensor [34]. In QCM biosensors, the resonant frequency is monitored to deduce changes in accumulated the surface mass concentration while sensing in gaseous or fluidic environments [35].

Thermometric biosensors use thermometry (measurement of temperature) to monitor biochemical interactions, using for instance a thermometer [28] or sensitive thermistors [36], as heat-producing biochemical absorption takes place. Thermometry has also been integrated with ELISA in a new method known as thermometric ELISA (TELISA) [37,38].

Optical biosensors measure an optical parameter such as absorption or refractive index, or optical emission such as fluorescence, luminescence, or Raman scattering [39-41]. Optical biosensors can provide highly sensitive, direct, real-time, and label-free detection of many biological and chemical substances [42]. Optical biosensors operate in label-based or label-free mode [42]. In label-based sensors, a fluorescent label is used, and the optical emission produced (or quenched) by a biochemical reaction is monitored. In label-free devices, the detected signal is generated directly by the interaction of the analyte with recognition elements immobilised on the transducer surface [42]. There are different varieties of optical biosensors, such as optrode fibre optics, evanescent-wave waveguides, flow immunosensors, and SPR.

Surface plasmon polaritons (SPPs) are propagating modes (optical surface waves) at the interface between a metal and a dielectric, and biosensors based on SPPs are among the most studied and used because of the small mode area and large surface fields associated with SPPs [43]. These features make SPPs very sensitive to changes in the refractive index near the metallic surface. However, propagation losses associated with the absorption of light in the metal is a limitation. This problem can be addressed using symmetric structures based on a thin metal layer to support long-range SPPs (LRSPPs) [44]. The existence of LRSPPs requires that both the upper and lower claddings have a similar refractive index. Sensing in an aqueous environment limits the choice of lower cladding materials, *e.g.*, to polymers such as Teflon or Cytop [20], which leads to limitations in the fabrication of devices, especially in terms of thermal excursions during processing. An alternative approach consists of using a one-dimensional photonic crystal (1DPC) [43]. A 1DPC can be used on one of the sides of a thin metal layer and they can mimic the optical properties of the medium on the other side over a limited wavelength range. The modes supported in such a structure are called Bloch LRSPPs.

A trend in (bio)chemical sensor research is to combine electrochemical techniques with surface plasmons. This combination is natural because both techniques involve reactions occurring on a metal surface. In such systems, surface plasmons can be used to probe the electrochemical double layer or to influence electrochemical reactions by involving energetic carriers that are naturally generated in the working electrode as surface plasmons are absorbed therein [30]. In this paper we introduce concepts related to optical sensing using SPPs and concepts related to electrochemical sensing in a tutorial-like fashion. We then review some representative literature on the integration of both techniques, aiming to realise multimodal sensors, or to investigate fundamentals by optically probing electrochemical activity or uncovering electrochemical effects due to optical absorption (pumping) of working electrodes. Section 2 introduces concepts of importance to SPPs on planar structures, Section 3 introduces concepts related to electrochemistry, and reviews non-exhaustively but representatively some literature on the integration of electrochemical techniques with surface plasmons. Section 4 gives a brief conclusion.

## 2. Surface Plasmons

### 2.1 Optics of Metals

Metals are traditionally used for circuitry and guiding microwaves or far-infrared EM waves. Metals can be considered perfect electric conductors (PEC) in these low-frequency applications because the conduction electrons cancel out the external field, effectively preventing wave propagation through the metal. Field penetration increases as frequencies increase toward the near-infrared and visible portions of the spectrum, which results in increased absorption and dissipation. Metals exhibit complex, frequency-dependent dielectric functions in this region of the spectrum. The metal permittivity can be described by various models. In the following, two well-known models, the Drude-Sommerfeld and Drude-Lorentz models are briefly explained.

The permittivity of metals at optical frequencies can be modelled simply as a plasma (free electron gas). Electromagnetic waves produce electron oscillations with respect to a lattice of positive ions. An oscillation can be described by its frequency, and interaction with the lattice by the collision frequency which is inversely proportional to the relaxation time of the free electrons. The expression for the dielectric function of a (non-magnetized) plasma is given by the Drude-Sommerfeld equation [45]:

$$\varepsilon(\omega) = 1 - \frac{\omega_p^2}{\omega^2 + i\gamma\omega} \qquad (1)$$

where $\omega_p$ is the plasma frequency, $\gamma$ is the collision frequency which is proportional to the inverse of the relaxation time of the electrons ($\tau = 1/\gamma$) [46], and $\varepsilon(\omega)$ is the relative permittivity. The plasma frequency is given by:

$$\omega_p = \sqrt{\frac{n_e e^2}{m^* \varepsilon_0}} \qquad (2)$$

in which $n_e$ is electron density, $e$ is the electron charge, $m^*$ is the effective electronic mass and $\varepsilon_0$ is the permittivity of free space. The plasma frequency occurs in the visible range for most metals. The frequency dependent dielectric function of a plasma has real and imaginary components:

$$Re\{\varepsilon(\omega)\} = 1 - \frac{\omega_p^2}{\omega^2+\gamma^2} \tag{3}$$

$$Im\{\varepsilon(\omega)\} = \frac{\gamma\omega_p^2}{\gamma(\omega^2+\gamma^2)} \tag{4}$$

The plasma (free electron gas) is metallic if $\omega_p > \omega$ and $\omega \gg \gamma$, as the permittivity is mostly real and negative.

For $\omega_p \leq \omega$, corresponding to short visible wavelengths and the ultraviolet region for most metals, the plasma is dielectric as the real part of permittivity becomes positive. Also in this region, the free electron gas model typically does not match the measured permittivity due to the onset of vertical (inter-band) electron transitions in the metal – in this case the Drude-Sommerfeld model needs a correction. Inter-band electron transitions increase the imaginary component of the permittivity dramatically [45]. Terms are added to the Drude-Sommerfeld model to improve the model of the dielectric function [47,48]:

$$\varepsilon(\omega) = \varepsilon_\infty - \frac{\omega_p^2}{\omega^2+i\gamma\omega} + \varepsilon_{int}(\omega) \tag{5}$$

where $\varepsilon_\infty$ is a material specific polarization correction factor and $\varepsilon_{int}(\omega)$ is a fitted permittivity which corrects for inter-band transitions, following [47,48]:

$$\varepsilon_{int}(\omega) = 1 - \frac{\tilde{\omega}_p^2}{(\omega_\circ^2-\omega^2)-i\gamma\omega} \tag{6}$$

where $\tilde{\omega}_p$ is analogous to $\omega_p$ in the Drude-Sommerfeld model. This model is called the Drude-Lorentz model.

Because of the complex optical response of metals, the optical permittivity is usually derived from experimental data [48,49]. Fig. 1 shows the real and imaginary parts of the optical permittivity of Au, following the Drude-Sommerfeld and Drude-Lorentz formulas, along with experimental data [50]. It is clear that the Drude-Sommerfeld model fits the measured data only at long wavelengths, $\lambda_0 > 700$ nm, whereas the Drude-Lorentz model captures the measurements accurately for $\lambda_0 > 500$ nm, along the long-wavelength side of the absorption feature which is caused by the onset of inter-band transitions in Au. Thus, Au behaves as a good optical metal at wavelengths longer than the absorption edge, $\lambda_0 > 700$ nm, into the infra-red, approaching a perfect electric conductor at long wavelengths beyond about 10 µm.

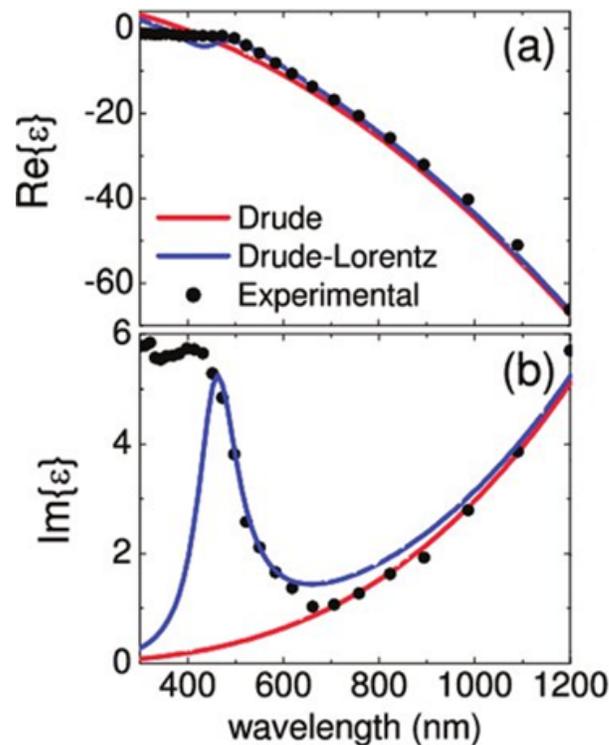

Fig. 1. (a) Real and (b) imaginary parts of the permittivity at optical wavelengths. Black dots correspond to experimental data. The red and blue curves are fits to the Drude-Sommerfeld and Drude-Lorentz formulas. Reprinted with permission from [50]. Copyright 2011 American Chemical Society.

## 2.2 Surface Plasmon Polaritons (SPPs)

Electromagnetic (EM) waves in the visible and near-infrared ranges can couple, under the right circumstances, to electron oscillations on the surface of a metal. The resulting excitation is termed a surface plasmon polariton (SPP), and it propagates as a surface wave along the metal-dielectric interface with fields that are maximum at the interface and decay exponentially (evanescently) away [45,50-52]. The two main groups of SPPs are propagating SPPs which exist on a plane metal surface, and localized SPPs which exist as resonant modes on a nanostructure such as a metal nanoparticle. Fig. 2 illustrates a propagating SPP along the interface of a metal and a dielectric.

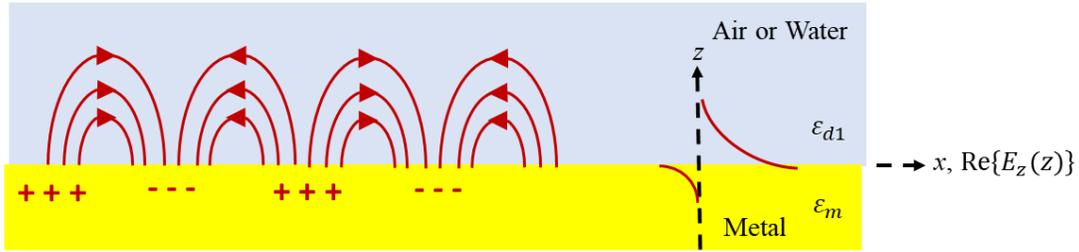

Fig. 2. Coupled excitation involving EM fields (curves) and an electron surface charge density wave (+ and -) forming a propagating SPP along a metal-dielectric interface.

As depicted in Fig. 2, the coupled excitation involves EM fields that induce a coherent oscillation of surface charges indicated by the "+" and "-" signs. The curved arrows indicate the associated electric fields. The SPP will propagate along the interface in the *x*-direction with a complex propagation constant $K_{spp}$ implying an exponential decay with propagation distance due to absorption.

According to the boundary conditions, the electric displacement field ($D_z$) must be continuous across the boundary, $\varepsilon_m E_{z,m} = \varepsilon_d E_{z,d}$, where $\varepsilon_d$ is the relative permittivity of the dielectric, $\varepsilon_m$ is the relative permittivity of the metal, and $E_z$ is the normal electrical field in both regions. Since $\varepsilon_m \neq \varepsilon_d$, there is a discontinuity at the surface in the electric field, resulting in surface charges at the interface. The real permittivity of a metal at optical frequencies is negative, while the permittivity of dielectrics is positive, resulting in a change in direction of the normal electric field across the interface, which enables the existence of an optical surface wave, in this case the SPP. TE (transverse-electric) polarized waves (*s*-polarized) have no electric field component perpendicular to the surface, so they cannot induce a polarization. Only TM (transverse-magnetic) polarized waves (*p*-polarized) can propagate.

Solving the vector wave equations for a planar interface between optically semi-infinite metal and dielectric regions (*i.e.*, a single interface) yields the dispersion equation for propagating SPPs as follows [54]:

$$K_{spp} = \frac{\omega}{c}\sqrt{\frac{\varepsilon_d \varepsilon_m}{\varepsilon_d + \varepsilon_m}} \tag{7}$$

in which $\omega$ is the optical angular frequency, and $c$ is the speed of light in free space. Based on the Drude-Sommerfeld model, $\varepsilon_m$ can be expressed as in Eq. (3). For visible and near-infrared light, $\gamma \ll \omega_p$, so Eq. (3) is often simplified to:

$$\varepsilon_m = 1 - \frac{\omega_p^2}{\omega^2} \tag{8}$$

The dispersion curve of the propagating SPP can be obtained by inserting Eq. (8) in Eq (7), yielding Fig. 3. SPPs behave like free-space photons at low frequencies, but their dispersion curves move increasingly to the right of the light line as frequency increases. At $\omega_L = \omega_p/\sqrt{2}$, $K_{spp}$ reaches an asymptotic limit. The dispersion curve bends to the right of the light line, so the SPP has greater momentum than incident light, which means it cannot be excited directly by light. Usually, a prism or a grating coupler is used to add momentum to the incident light. By using these mechanisms, the momentum mismatch is compensated, allowing light to couple into propagating SPPs that travel along the dielectric-metal interface.

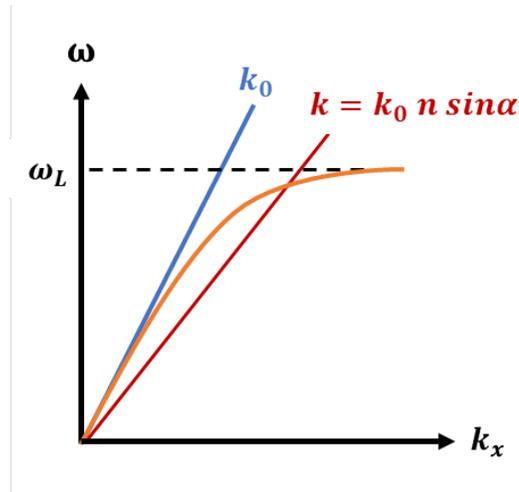

Fig. 3. Dispersion plot of a propagating SPP showing the problem of momentum mismatch between illuminating light (blue line) and the SPP (orange curve). It is necessary to overcome momentum mismatch in order to couple the incident light to the SPP, because the SPP is on the right side of the light line, having higher momentum ($K_{spp}$) than a free space photon ($k_0$) of the same frequency. In order to provide additional momentum, one can use evanescent wave coupling in total internal reflection from a prism, $k = k_0 n \sin \alpha$ (red line).

Fig. 4(a) illustrates the 1D normal electric field profile (real part) of the single-interface SPP on a semi-infinite metal of relative permittivity $\varepsilon_m$ bounded by a semi-infinite dielectric of relative permittivity $\varepsilon_{d1}$ [47]. The field is observed to peak at the interface and decay exponentially on each side. A single-interface SPP propagates along the surface until its energy is dissipated via absorption in the metal or scattered. This leads to a propagation length typically of the order of 10 to 100 µm at visible and near-infrared wavelengths. The metals most useful to support SPPs over this wavelength range are Al and Ag, with Au and Cu being limited primarily to the near-infrared [56].

Fig. 4(b) illustrates a finite thickness metal slab with dielectrics on both sides. Here, the two SPPs supported by the top and bottom interfaces couple through the metal to become coupled modes (sometimes referred to as mixed modes or supermodes). They are described by symmetric ($s_b$) or asymmetric ($a_b$) transverse electric field profiles as sketched.

An interesting condition occurs for the modes on the metal slab if the permittivity of the surrounding dielectrics (lower-cladding and upper-cladding) are equal ($\varepsilon_{d1} = \varepsilon_{d2}$), and both claddings are lossless (dielectric) [44]. The symmetric mode has a decreasing attenuation with a decreasing metal thickness as the electric field within the metal will be increasingly expelled. Conversely, the asymmetric mode has increased confinement in the metal with decreasing thickness, and thus increasing attenuation. The symmetric mode is known as a long-range surface plasmon polariton (LRSPP). LRSPP modes have lower attenuation (at least one to two orders of magnitude) than their single interface and asymmetric counterparts. The lower attenuation allows for longer propagation lengths and much longer interaction lengths in sensor applications.

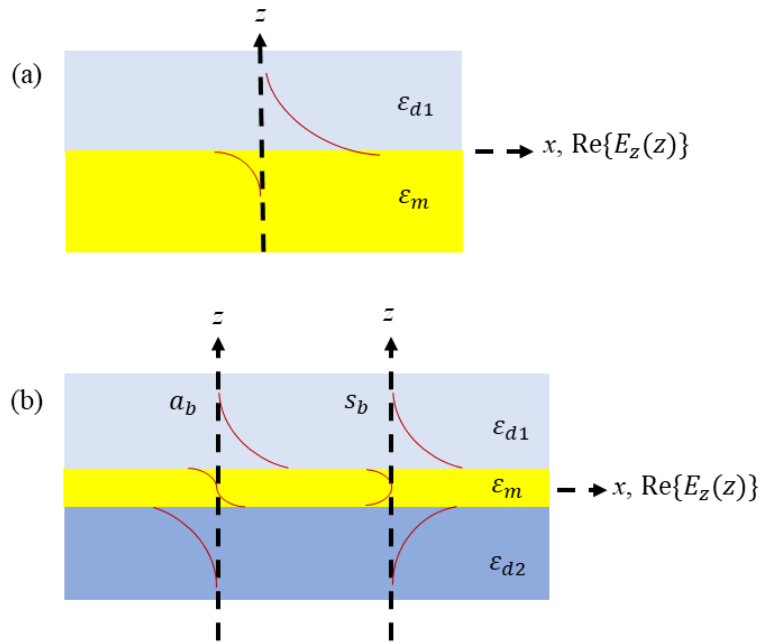

Fig. 4. Sketch of 1D normal electric field profiles (real part), Re{$E_z(z)$}, for (a) the single-interface SPP, and (b) two coupled SPP modes on a thin metal film, $a_b$ and $s_b$. The propagation direction may be taken along the $y$ axis.

The metal slab sketched in Fig. 4(b) can be limited to a width *w*, forming a metal stripe, as depicted in Fig. 5, which introduces lateral (horizontal) confinement. This limitation dramatically changes the modal solutions, adding modes to the system that are symmetric or asymmetric about the *z*-axis. This leads to four fundamental modes, including one that is symmetric about both the *x* and *z* axes, referred to as the $ss_b^\circ$ mode [57]. If the permittivity of the lower cladding equals that of the upper cladding, the $ss_b^\circ$ mode exhibits vanishing attenuation as the metal stripe vanishes (*w*, *t* → 0) and is referred to as the LRSPP mode of the metal stripe waveguide [57]. Furthermore, the symmetric LRSPP profile is amenable to efficient end-fire excitation by an incident Gaussian beam or by butt-coupling to an optical fibre [44]. The metal stripe waveguide enables integrated optical structures such as straight waveguides, bends, splitters, and Mach-Zehnder interferometers, useful for sensing applications [58].

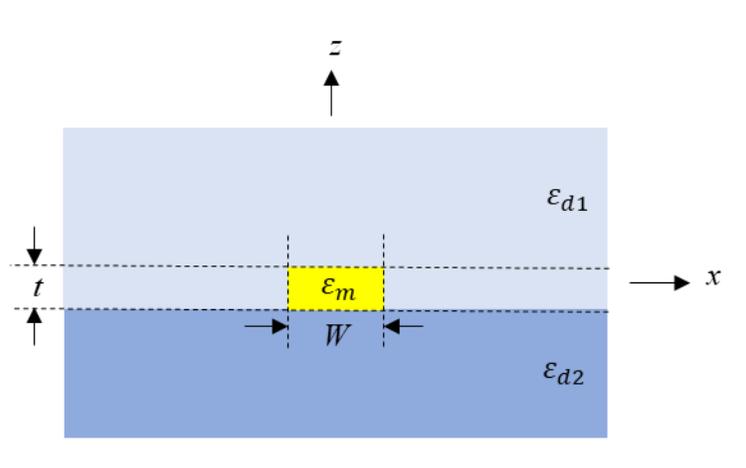

Fig. 5. Metal stripe of thickness *t* and width *w* bounded by dielectrics. The substrate and cladding are generally not the same, but $\varepsilon_{d1} = \varepsilon_{d2}$ is required for LRSPP propagation. The propagation direction is along the *y* axis.

The lateral confinement changes the modal solutions dramatically, and modal analysis requires the use of numerical methods. Numerical methods for modal calculations include the method of lines (MoL) [59], the finite element method (FEM) [60] or the finite difference method (FDM) [61].

For waveguide structures shown in Figs. 4 and 5, the SPP mode propagates in the $y$ direction with a complex propagation constant, $\gamma = \alpha + j\beta$, where $\alpha$ is the attenuation constant and $\beta$ is the phase constant ($e^{+j\omega t}$ time harmonic form implied). The mode power attenuation (MPA) in dB/m computed from attenuation constant $\alpha$ is [46]:

$$MPA = 20\alpha \log_{10} e \tag{9}$$

The mode power is reduced by a factor of 1/e in a distance from the launch point defined as the propagation length of the mode ($L_e$) [46]:

$$L_e = \frac{1}{2\alpha} \tag{10}$$

## 2.3 Excitation of Surface Plasmon Polaritons

To excite SPPs, both energy and momentum (phase matching) conservation must be fulfilled. Phase matching requires that the propagation constant of the input light ($k = \omega/c$) be equal to the propagation constant of the SPP ($K_{spp}$). Fig. 3 shows that $K_{spp}$ is always larger than $k$ for single-interface SPPs, so the direct excitation of SPPs by light is not possible.

The most common technique to excite single-interface SPPs is via prism coupling where the attenuated total reflection (ATR) condition is indicative of SPP excitation at the metal dielectric interface. The Otto configuration, sketched in Fig. 6(a), is one prism-coupling method used to excite SPPs [67], where a dielectric gap exists between the prism and the metal surface. The Kretschmann–Raether configuration, sketched in Fig. 6(b), is another technique for exciting SPPs, and is more convenient as the metal film is deposited directly onto the base of the prism [67]. A prism-coupled system based on the Kretschmann-Raether configuration is the most common method of exciting SPPs in SPR sensors [64-69].

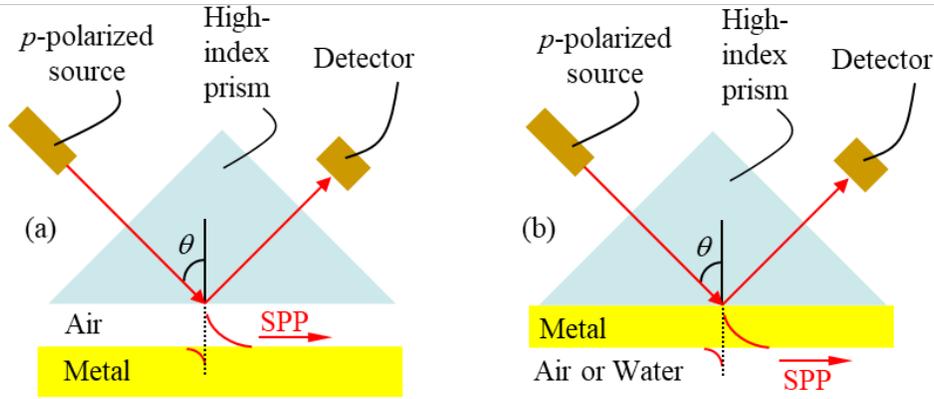

Fig. 6. Excitation of SPPs using (a) Otto and (b) Kretschmann-Raether configurations.

In the Kretschmann geometry, the momentum matching condition for the excitation of SPPs on the metal film is described as:

$$k_x = \left(\frac{2\pi}{\lambda}\right) n_p \sin\theta = Re\{K_{spp}\} \qquad (11)$$

where $k_x$ is the in-plane wavenumber of the incident light at incident angle $\theta$ which can couple into SPPs, $n_p$ is the refractive index of the prism, and $\lambda$ is the free-space wavelength.

Prism coupling has been employed extensively to excite SPPs at the metal liquid interface, and applied to a large number of biodetection problems, *e.g.*, in bulk RI sensing [71], environmental bio-sensing of pesticides [72], effects of toxins on cells [73], dual-parameter sensing [74], protein bio-sensing [75], and monitoring cellular micro-motion within fibroblast cells [76]. Moreover, this method has been used in detection of *e.coli* bacteria [77], foodborne pathogens [78], Bovine Serum Albumin (BSA) [79], cardiac troponin [80], Benzo[a]pyrene (BaP) [81], and pituitary hormones [82].

LRSPP excitation can be done either by a prism coupled system based on attenuated total reflection (ATR), or as discussed in the previous sub-section in an end-fire coupling scheme using a TM-polarized Gaussian beam or a polarization-maintaining single mode fibre (PM-SMF) (butt-coupling) [83,84]. In a prism-coupled system, the incident angle can be changed beyond the critical angle and the reflected power is monitored using a photodetector as shown in Fig. 7.

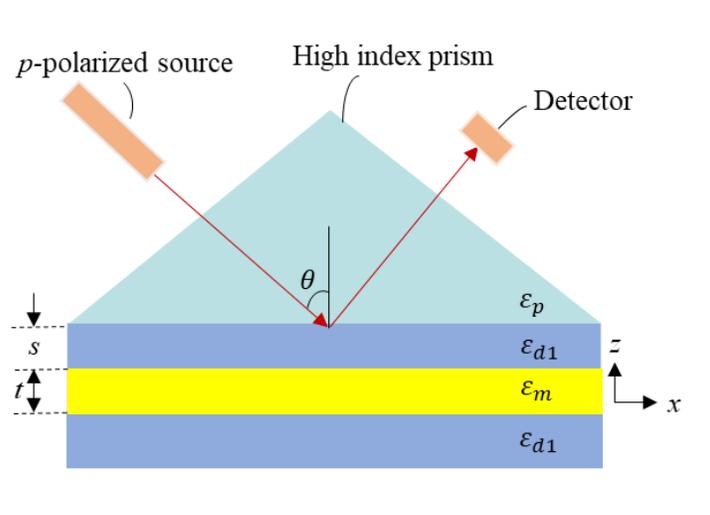

Fig. 7. Prism coupling technique for exciting LRSPPs; $\varepsilon_p$ is the relative permittivity of the prism (= $n_p^2$) and $\varepsilon_{d1}$ is the relative permittivity of the dielectrics bounding the metal, the outer one being the sensing solution.

The in-plane wavenumber of the incident beam is equal to the wavenumber of the propagating LRSPP mode at a certain angle, and a drop in the reflected power is observed (similar to the Kretschmann-Raether configuration for single-interface SPPs). In this configuration a thin metal slab is bounded by two dielectric (insulator) layers forming an insulator-metal-insulator (IMI) structure that can support LRSPPs. For sensing applications, the dielectric layer separating the metal film from the prism should have a relative permittivity of $\varepsilon_{d1}$ matching that of the aqueous sensing fluid on the other side of the film. Although prism coupling is simple, it can be somewhat bulky and often impractical.

Another method of exciting SPPs and LRSPPs is with grating couplers. Fig. 8 shows a grating, formed of either periodic bumps or trenches, of period $\Lambda$. Grating couplers must be designed to satisfy the momentum conservation condition:

$$\Lambda = m\lambda(n_{eff} - n_c \sin\theta) \qquad (12)$$

where m is the (integer) order of the grating, $n_{eff}$ is the average effective index of the SPP propagating along the waveguide with the grating, $\lambda$ is the free-space wavelength, $n_c$ is the refractive index of the medium from which the beam is incident, and $\theta$ is the angle of incidence.

Fig. 8 illustrates a metal grating, but the grating can be patterned in another material such as a dielectric. Grating couplers work by scattering incoming light of propagation constant $k$ at incident angle $\theta$ into SPPs. The component of the scattered light which matches the propagation constant of the SPP ($K_{spp}$) will couple. Using Eq. (12), the phase matching condition (momentum conservation) is written:

$$K_{spp} = k\sin\theta \pm m\frac{2\pi}{\Lambda} \qquad (13)$$

Grating couplers are easy to align and excite but the coupling efficiency is limited due to the nature of the structure. This approach has been implemented with metal stripes in Cytop [85], and with metal stripes on a truncated photonic crystal to excite Bloch-LRSPPs [43,86].

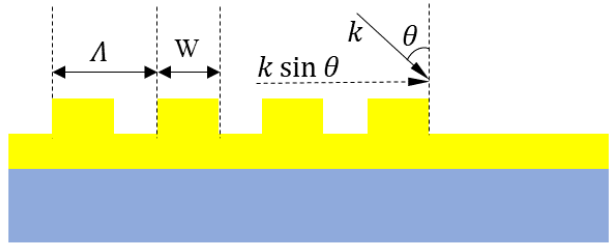

Fig. 8. Geometry of a grating coupler for SPPs as a step-in-height metal pattern.

As mentioned previously, end-fire coupling is a particularly apt scheme for exciting LRSPPs on the metal stripe waveguide (*cf.* Fig. 5(a)), because the mode fields of the LRSPP, Fig.

5(b), overlap very well with those of a PM-SMF or with an incident Gaussian beam [44]. In such an excitation scheme, the coupling efficiency can be obtained by estimating the overlap factor $C$ of the LRSPP mode with the source field:

$$C = \frac{\iint E_{z1} \cdot E_{z2}^* \, dA}{\sqrt{(\iint E_{z1} \cdot E_{z1}^* \, dA)(\iint E_{z2} \cdot E_{z2}^* \, dA)}} \tag{14}$$

which is computed from the spatial distribution of the main transverse electric field component of the LRSPP, $E_{z1}$, and of the source field, $E_{z2}$. The mode power coupling efficiency is given by $|C|^2$ if there is no discontinuity in the materials at the coupling plane. The mode power coupling loss ($C_{pl}$) is expressed in dB:

$$C_{pl} = -20 \, \log_{10} |C| \tag{15}$$

End-fire coupling to a PM-SMF yields coupling efficiencies of ~90% or greater to LRSPPs on a metal stripe and has been exploited extensively to interrogate biosensors based on metal stripe waveguides [20].

**2.4 Bloch long-range surface plasmon polaritons (Bloch LRSPPs)**

In biosensing applications based on metal stripe waveguides (Fig. 5), the lower cladding material must be selected such that its refractive index is close to that of the sensing fluid which also acts as the upper cladding. This is needed to ensure that the LRSPP is supported, because a symmetric dielectric environment is required near the metal stripe, and to ensure that any microfluidic channels etched into the cladding to define the sensing region is optically non-invasive once filed with the sensing fluid. Most sensing fluids in biosensor applications are aqueous in nature - clean buffers carrying analyte or patient samples (urine, blood) diluted in buffer, for example. Thus, the choice of bottom cladding or substrate should be limited to materials that have a refractive index close to water, such as Cytop [87,20] and Teflon [88].

An alternative approach consists of using a truncated one-dimensional photonic crystal (1DPC) as a substrate [43]. This approach has the advantage of material flexibility because several inorganic materials can be used to implement the stack. A 1DPC can be used as a lower cladding because it can be designed to mimic the optical properties of the medium on the other side of the metal stripe or slab over a range of wavenumber and wavelength. A mode supported in this structure is termed a Bloch LRSPP.

Fig. 9(a) shows such a structure. The 1DPC is formed on a Si substrate and covered by an optically semi-infinite Cytop layer or by the sensing medium. The 1DPC depicted was formed as a $SiO_2/Ta_2O_5$ multilayer stack, designed to tailor the field decay into the stack and minimize losses from light tunnelling into the Si substrate [43].

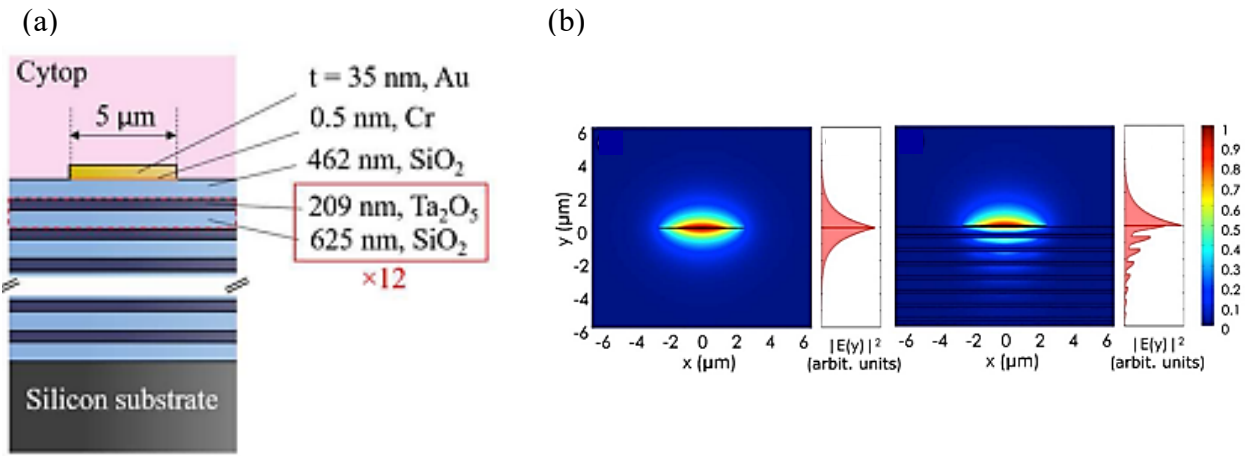

Fig. 9. (a) Au stripe on a truncated 1DPC covered by Cytop, supporting Bloch LRSPPs. (b) Profiles of the squared magnitude of the perpendicular electric field of the LRSPP on a Au stripe buried in Cytop (left panels), and of the Bloch LRSPP on a Au stripe on a truncated 1DPC covered in Cytop (right panels). Reprinted with permission from [43]. Copyright 2017 American Chemical Society.

The distribution of the squared magnitude of the perpendicular electric field component of the LRSPP in the corresponding Cytop embedded structure is shown on the left panels of Fig. 9(b), and of the Bloch LRSPP on the truncated 1DPC on the right panels of Fig. 9(b). The field distributions extend similarly into the claddings, with the exception that the Bloch LRSPP has an oscillatory character as it decays into the IDPC. Both modes present essentially the same surface sensitivities as they overlap similarly with a thin adlayer on the metal stripe [89,90].

Fig. 10 illustrates schematically an excitation scheme for Bloch LRSPPs consisting of illuminating a Au grating coupler on a Au stripe with a *p*-polarized Gaussian beam launched using an aligned PM-SMF. Such an excitation scheme has the advantage of not requiring a high-quality input facet, as well as simpler optical alignments in comparison to butt-coupling. However, the coupling efficiency is typically lower than end-fire coupling in such a scheme. The grating consists of rectangular ridges of dimensions $W$ and $H$ disposed over a period of $\Lambda$. The grating was modeled in the frequency domain with the 2D FEM using software available commercially, and the coupling efficiency was calculated by capturing the power carried by the Bloch LRSPP far away from the grating region (150 µm to the left) to eliminate the effects of spatial transients, as described in [91]. A coupling efficiency of 16% was deduced at $\lambda_0 = 1310$ nm over about 70 nm of optical bandwidth [43].

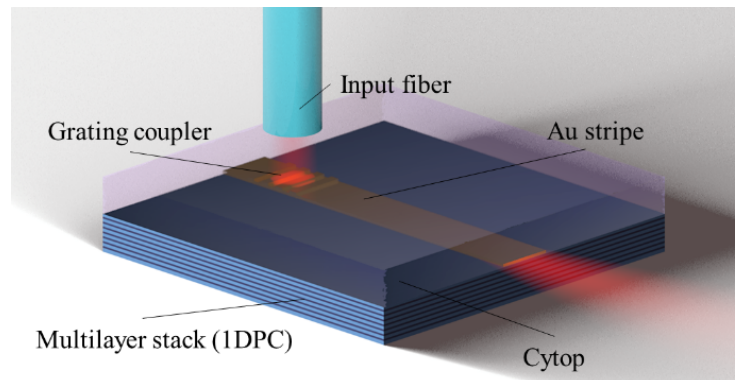

Fig. 10. 3D sketch of the metal stripe waveguide with a grating coupler on a 1DPC, illustrating this excitation arrangement. Reprinted with permission from [43]. Copyright 2017 American Chemical Society.

**2.5 Optical Interrogation of Surface Plasmon Biosensors**

Four primary optical interrogation methods can be applied to a surface plasmon sensor [92-94], depending in part on the excitation scheme used - prism, grating, or end-fire coupling. Firstly, phase interrogation, where a phase shift in the collected light is measured indicating a change in refractive index, with the wavelength and excitation conditions remaining constant – this scheme must be incorporated into an interferometer to convert phase changes to changes in intensity. Secondly, angular interrogation, where a single wavelength (monochromatic) laser is used and the shift in angle of SPP coupling is measured indicating refractive index changes – this scheme commonly termed SPR requires a prism or grating, and fundamentally also interrogates phase, but by varying the coupling conditions. Thirdly, spectral interrogation uses a broadband source with a spectrograph or a tunable laser with a power sensor, measuring a shift in the resonant or coupling wavelength in reflection or transmission to examine the change in refractive index – this scheme is broadly applicable using prism, grating or end-fire coupling. Fourthly, intensity interrogation measures the change in transmitted or reflected light intensity, while maintaining other excitation conditions constant.

Structures for sensing applications can be characterized by parameters such as bulk sensitivity, surface sensitivity, and figure of merit (FOM). The bulk sensitivity is defined as a change in measurand, *e.g.*, a coupling or resonant wavelength $\lambda_{res}$ *vs*. the change in refractive of the bounding medium (*e.g.*, $\partial \lambda_{res}/\partial n_c$). By changing the refractive index of the bounding medium from a nominal value to another, step by step, one can determine the bulk sensitivity [92].

The metal surface in a biosensor is in contact with a sensing solution (bounding medium) and functionalized chemically to react selectively with the target analyte. As analyte attaches, an adlayer of refractive index higher than the sensing solution forms, causing a shift in measurand. The surface sensitivity can be defined as the shift in the measurand, *e.g.*, a resonance wavelength

$\lambda_{res}$, as a function of adlayer thickness $a$ forming at the metallic/solution interface (*e.g.*, $\partial \lambda_{res}/\partial a$).

Conventional SPR biosensors based on the Kretschmann-Raether configuration use a glass prism to excite SPPs. An immobilized bio-recognition element is coated on the metal surface as shown in Fig. 11. Plasmon excitation occurs at a specific angle termed the SPR angle. At this angle, the incident light is coupled to SPPs which leads to a decrease in the intensity of the reflected beam. The refractive index of the sensing solution and the presence of an adlayer at the metal/solution interface determine the SPR angle. Biochemical reactions change the thickness of the adlayer which changes the SPR angle required to maintain excitation of SPPs. Changing the refractive index of the dielectric medium on the other side of the metal film, or forming an adlayer thereon, will result in significant changes in the SPP coupling angle (Eq. (11), as $K_{spp}$ is altered). Plotting the intensity *vs*. the incidence angle over time produces a sensorgram from which binding kinetics can be extracted [68,69].

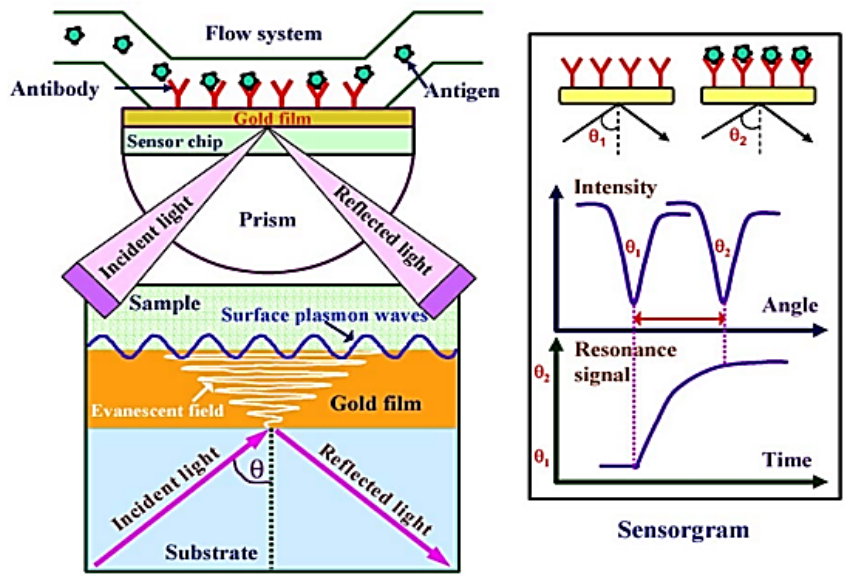

Fig. 11. Schematic view of the SPR immunoassay technique. Reprinted in part with permission from [69]. Reprinted from Publication Shankaran, D. R.; Gobi, V.; Miura, N. Recent advancements in surface plasmon resonance immunosensors for detection of small molecules of biomedical, food and environmental interest. Sens. Actuators B Chem, **2007**, 121, 158–177. Copyright 2007, with permission from Elsevier.

Wavelength interrogation is also possible, where the excitation of propagating SPPs is revealed as the appearance of a dip in the reflectance spectrum for a fixed angle of incidence. The spectral width of this dip is typically ~50 nm full width at half-maximum (FWHM) for Au films [70], and it is accompanied by an apparent phase change near the minimum of the resonance for the reflected light. SPR biosensors conventionally exploit single-interface SPPs which have a high attenuation and broad resonance conditions.

## 3. Electrochemical surface plasmon sensors

A recent trend in (bio)chemical sensors is to combine electrochemical techniques with surface plasmon sensors. Such a merger is natural because both techniques exploit reactions occurring on a metal surface, often Au due to its inertness, in both types of systems. In such systems, surface plasmons can be used to probe the electrochemical double layer or to influence electrochemical reactions by involving energetic carriers that are naturally generated in the working electrode as surface plasmons are absorbed therein. For instance, electrochemical surface plasmon resonance (EC-SPR) probes faradaic processes by monitoring the change in refractive index that happens with the change in redox state at the electrode surface in electrochemical Kretschmann SPR systems [95]. EC-SPR is an important application of SPR to study local electrochemical reactions on the surface of the electrode. We discuss briefly cyclic voltammetry before reviewing the literature on such systems.

## 3.1 Cyclic Voltammetry

Cyclic voltammetry (CV) is a powerful electrochemical technique that investigates the reduction and oxidation processes of redox species [96]. The traces in Fig. 13 are termed voltammograms or cyclic voltammograms [97]. Here, the *x*-axis corresponds to the potential applied to the system (V), and the *y*-axis represents the response (measurand), which is the resulting current (C). CV data are commonly reported using two conventions, but the sign convention used to obtain and plot the data is rarely stated. As shown in Fig. 12, the potential axis clarifies the convention.

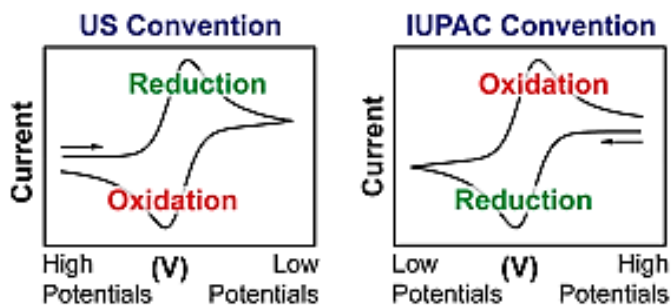

Fig. 12. Schematic voltammograms. To report CV data, US and IUPAC conventions are commonly used. The data reported in the two conventions appear rotated by $180°$. Reprinted with permission from [97]. Copyright 2018 American Chemical Society.

During a CV experiment, the potential is varied linearly at a rate of a few millivolts per second. This parameter is called the scan rate and it is one of the most important parameters in cyclic voltammetry. Fig. 13(a) shows the relation between time and applied potential with the potential plotted on the *x*-axis, commensurate with the corresponding voltage voltammogram of Fig. 13(b). In Fig. 13(a), the potential is swept positively in the forward scan from the starting potential $E_1$ to the switching potential $E_2$. This is termed the anodic trace. The scan direction is then reversed, and the potential is swept negatively back to $E_1$, called the cathodic trace.

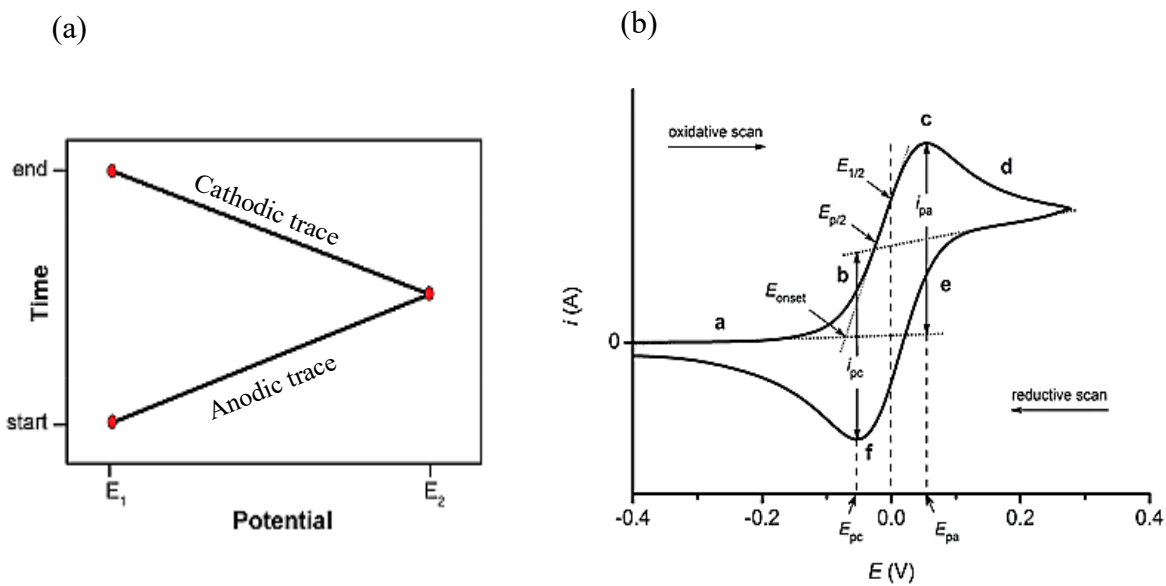

Fig. 7 (a) Applied potential as a function of time for a generic cyclic voltammetry experiment, with the initial, switching, and end potentials identified. (b) Voltammogram of a reversible reaction (IUPAC convention). Reprinted with permission from [97]. Copyright 2018 American Chemical Society.

The Nernst equation holds at equilibrium between the oxidized and reduced species [97]:

$$E = E° + \frac{RT}{nF} \ln \frac{([OX])}{([Red])} = E° + 2.3026 \log_{10} \frac{([OX])}{([Red])} \tag{16}$$

The Nernst equation defines the potential of an electrochemical cell ($E$) as the standard potential of a species ($E°$) and the relative activities of the oxidized and reduced species in the system at equilibrium. [Ox] and [Red] denote the concentration of the oxidized and reduced species, $F$ is Faraday's constant, $R$ is the universal gas constant, $n$ is the number of electrons transferred in the reaction, and $T$ is the temperature.

The formal potential is specific to the experimental conditions and is often determined by $E_{1/2}$ (average potential of points f and c in Fig. 14(b)). In response to a change in the concentration of a species in solution or a change in electrode potential, the Nernst equation provides a powerful way to predict how a system will respond. When $E = E° = E_{1/2}$ is applied to a system, the Nernst equation predicts that the oxidized form will be reduced until the concentrations of the reduced and oxidized species are equal, and equilibrium is achieved. When the potential is scanned during the CV experiment, the concentration of the species in solution near the electrode changes over time based on the Nernst equation.

Reduction locally at the electrode occurs when a solution of oxidized species is scanned to negative potentials (cathodic direction) from the rightmost point in Fig. 13(b), resulting in a current and depletion of oxidized species at the electrode surface. At point f of Fig. 13(b), where the peak cathodic current ($i_{pc}$) is observed, the current is determined by the delivery of additional oxidized species via diffusion from the bulk solution. During the scan, the diffusion layer continues to grow at the surface of the electrode containing the reduced species. As a result, mass transport of the oxidized species slows. Therefore, upon scanning to more negative potentials, the diffusion rate of the oxidized species from the bulk solution to the electrode surface slows down, resulting in a decrease in current.

When the minimum potential is reached, the scan direction is reversed, and the potential is scanned in the positive (anodic) direction. The concentration of reduced species at the electrode surface decreases, while the concentration of oxidized species at the electrode surface increases, satisfying the Nernst equation. As the applied potential becomes more positive, the reduced species present at the electrode surface is oxidized. Following the Nernst equation, at $E = E_{1/2}$ the concentrations of oxidized and reduced species at the electrode surface are equal. These concentrations occur at the two points corresponding to the middle potential between the two observed peaks (oxidation c. and reduction f), which provides a straightforward way to estimate $E°$ for a reversible 1-electron transfer reaction. Due to the diffusion of redox species to and from the electrode, the two peaks are separated.

When the electrode reaction is controlled by diffusion, the plot of peak currents (anodic and cathodic), $i_p$, vs. the square root of the scan rate ($\vartheta$) must satisfy the Randle-Sevcik equation [83]:

$$i_p = 0.4463 \, nFAC \left(\frac{n F \vartheta D}{RT}\right)^{1/2} \qquad (17)$$

where $i_p$ is the peak current, $A$ is the electrode area, $D$ is the diffusion coefficient, $C$ is the concentration of the redox species, and $\vartheta$ is the scan rate.

## 3.2 Electrochemical Surface Plasmon Resonance (EC-SPR)

An interesting study of SPR to probe electrochemical reactions was reported by Wang *et al*. [98]. In this work, EC-SPR led to a new way to measure convolution voltammetry directly without the need for numerical integration of the electrochemical current response. With convolutional voltammetry, it is possible to determine diffusion constants, bulk concentrations, and the number of electrons transferred between electroactive species.

The EC-SPR signal is proportional to the time convolution of the electrochemical current density measured by conventional electrochemical methods [98]:

$$\Delta\theta(t) = B(\alpha_R D_R^{-1/2} - \alpha_0 D_0^{-1/2})(nF\pi^{\frac{1}{2}})^{-1} \int_0^t i(t')(t-t')^{-1/2} dt' \quad (18)$$

where $\Delta\theta = \theta(t) - \theta_0$ measures the changes in SPR resonance (coupling) angle, the constant $B$ represents the sensitivity of the SPR angle to changes in the bulk index of refraction (bulk sensitivity), which can be calibrated for a given SPR setup and reaction species, $D_0$ and $D_R$ are the diffusion coefficients of the reaction species, and $\theta_0 = B(\alpha_0 C_0^0 + \alpha_R C_R^0)$ is the SPR angle at $t = 0$.

The experiments reported used a prism-based SPR imaging setup as illustrated in Fig. 14(a). This experiment utilized a BK7 triangle prism with a collimated red LED (wavelength 670 nm) as the light source and a high-speed CCD camera as the detector. As the SPR sensing surface, a gold-coated microscope coverslip was placed on the prism using index matching fluid. Using the top opening of the electrochemical cell, a platinum wire counter electrode and an Ag/AgCl reference electrode were inserted into the electrolyte.

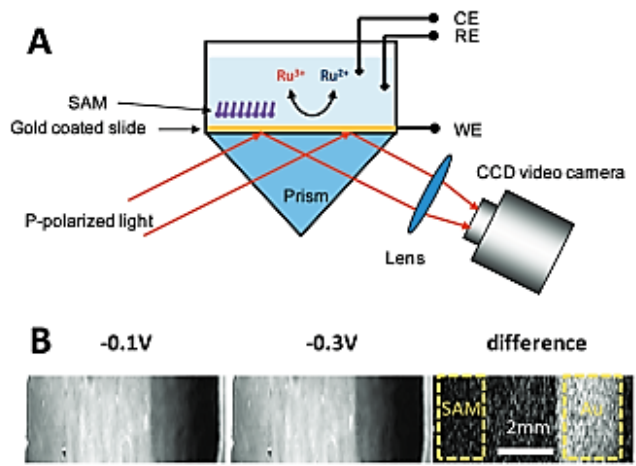

Fig. 8. (a) Illustration of the SPR setup. (b) SPR images in 11.78 mM Ru $(NH_3)_6^{3+}$ dissolved in phosphate buffer (0.5 M pH 7) at -0.1 and -0.3 V (*vs*. Ag/AgCl reference electrode) and their differences. Bare Au and SAM-coated areas used for data extraction and processing are indicated by yellow squares. Reprinted with permission from [98]. Copyright 2010 American Chemical Society.

In Fig. 15(a), the cyclic voltammograms are plotted for three scan rates, 0.01, 0.1, and 1.0 V/s, demonstrating that the peak current increases by 10 times when the scan rate is increased from 0.01 to 1.0 V/s. SPR voltammograms simultaneously recorded at different scan rates plotted in Fig. 15(b) show that the SPR response is not strongly dependent on the scan rate, as expected from theory (*cf*. Eq. (18)). Fig. 15(c) plots the computed SPR voltammograms using the measured current responses of Fig. 15(a), yielding reasonable agreement with the directly measured SPR voltammograms plotted in Fig. 15(b).

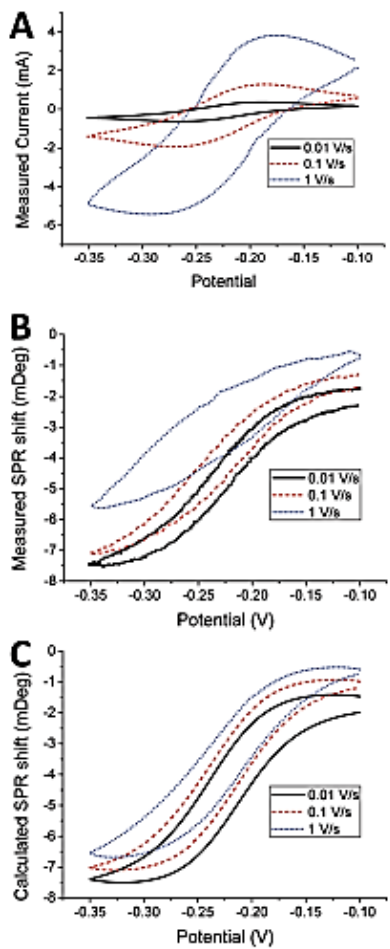

Fig. 9. (a) Measured current voltammograms. (b) Measured SPR voltammograms. (c) SPR voltammograms calculated using Eq. (18) and the results of panel (a) as inputs. The electrolyte was 3 mM Ru $(NH_3)_6^{3+}$ in phosphate buffer and the electrode/SPR surface was bare gold. Reprinted with permission from [98]. Copyright 2010 American Chemical Society.

Another study involving EC-SPR investigated how electrochemical reactions affect the metal/liquid interface to shift the reflection minimum. A prism covered by a silver film was utilized in an EC-SPR configuration with perchlorate and halide electrolytes as an optical probe of electrochemical reactions [99]. As the electrode potential becomes more positive, the resonance shifts to a smaller wavevector at a fixed wavelength, as shown in Fig. 16. Changes in electron density at the metal surface, ion adsorption, and changes in optical properties of the ionic double layer are some of the factors contributing to this shift.

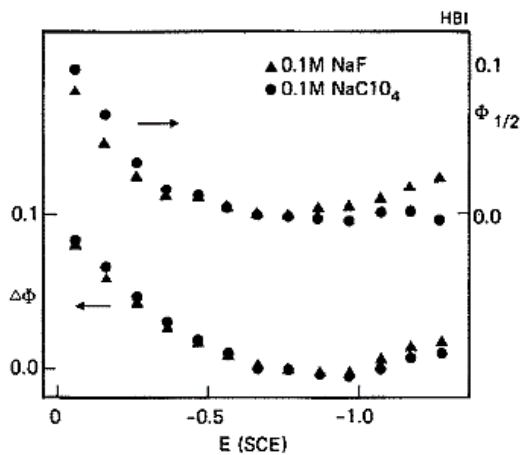

Fig. 10. Shift in SPR peak position and width vs. potential for a silver coated prism in 0.1 M NaF and 0.1 M NaClO$_4$. The shifts are relative to the position and width at -0.76 V [99]. Reprinted from Publication Gordon, J. G.; Ernest, S. Surface plasmons as a probe of the electrochemical interface. Surf. Sci., 1980, 101, 499-506. Copyright 1980, with permission from Elsevier.

EC-SPR was also demonstrated to be a highly effective tool for detecting and measuring intermediates and is being explored as a potential tool for studying reaction kinetics. The semiquinone radical anion BQ$^{·-}$ was detected in the hydroquinone-benzoquinone ion system by producing a large negative shift in a flow-through EC-SPR system, as shown in Fig. 17 [100]. The study demonstrates that flow-through EC-SPR can be used to monitor not only surface interactions but also chemical reactions in solution.

During potentiostatic oxydoreduction, an EC-SPR biosensor with an absorptive redox mediator film was used to detect reversible refractive index changes in the film. Using the organic dye Methylene Blue (MB) as an electroactive label, this work examined the theoretical and experimental foundations of EC-SPR sensing. The electrochemical activation of the SPR response is dependent on the local MB concentration and can be used to design highly sensitive and selective biosensing systems [101].

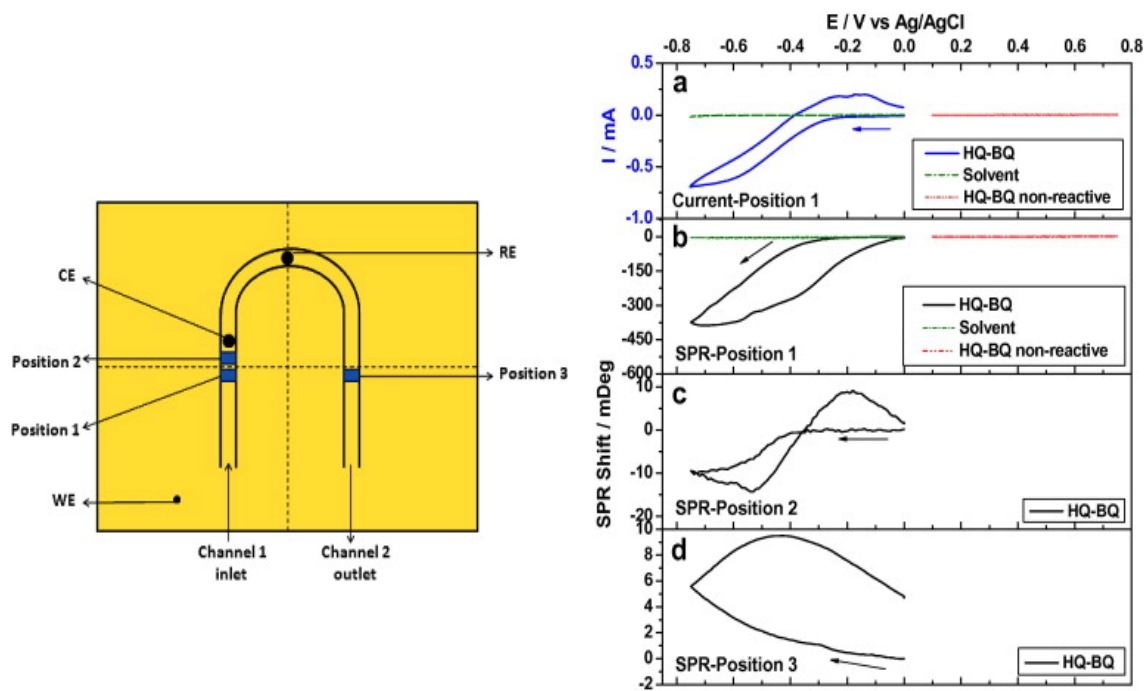

Fig. 11. Left panel: Gold-coated sensing chip with flow channels and electrodes shown schematically in top view. The two channels are connected in a U-shape. The width, height, and length of each channel are respectively 2.0, 0.5 and 5 mm. The working electrode (WE) is connected to the sensing chip via a contact pin. The reference electrode (RE) and counter electrode (CE) are in the flow channel as shown on the sketch. The solution enters the cell through channel 1 and exits through channel 2. Position 1 is located on the WE, while positions 2 and 3 are downstream. For a 100 µL/min flow rate, it takes 0.1 s and 5 s for a sample to flow from position 1 to positions 2 and 3, respectively. Right panel: Characterization of BQ$^{•-}$ with CV-SPR at positions 1, 2 and 3 of the sensing chip. (a) Current of the HQ–BQ and of the solvent (0.1 M Bu$_4$NPF$_6$ in the acetonitrile), and of HQ–BQ in a non-reactive potential window. (b) SPR signal of the HQ–BQ and of the solvent (0.1 M Bu$_4$NPF$_6$ in the acetonitrile), and of HQ–BQ in a non-reactive potential window at position 1. (c) SPR signal of the HQ–BQ at position 2. (d) SPR signal

of the HQ–BQ at position 3. Potential scan rate: 0.1 V/s. Flow rate 100 μL/min. Reprinted from Publication Huang, X.; Wang, S.; Shan, X.; Chang, X.; Tao, N. Flow-through electrochemical surface plasmon resonance: detection of intermediate reaction products. J. Electroanal. Chem., 2010, 649, 37-41. Copyright 2010, with permission from Elsevier.

**3.3 Energetic (Hot) Charge Carriers in EC-SPR**

Energetic (hot) charge carriers refer to either photoexcited holes or electrons that exist in non-equilibrium high-kinetic-energy states in photoactive materials, *e.g.*, metals and semiconductors, after being exposed to photons [103,104]. Photoexcited, non-equilibrium hot carriers in metallic structures could lead to bandgap-free photodetection and selective photocatalysis [105]. However, hot carrier devices must be significantly improved to meet practical application requirements. A promising pathway to increase the efficiency of these systems is to involve the excitation of SPPs [105].

The energy in an SPP is dissipated as free-space radiation (radiative loss) through scattering, or as absorption (nonradiative decay) in the metal. The absorption of SPPs produces energetic carriers - electrons and holes - in the metal that are not in thermal equilibrium with the lattice. These non-equilibrium hot carriers enable energy-harvesting applications in photovoltaics, photodetection, photon up-conversion and photocatalysis [103-108].

Recent attention has focused on studying the role played by energetic carriers created by SPP absorption in electrochemical reactions [30]. Plasmonic electrocatalysis holds promise for opening new reaction pathways inaccessible thermally, or for improving the efficiency of chemical processes. However, the underlying mechanisms of hot carrier transfer in photochemical processes remain mysterious, particularly those involving hot holes. Using photoelectrochemistry, hot holes and hot electrons can be localized on photoanodes and photocathodes, allowing investigation of hole-transfer and electron-transfer dynamics in oxidation and reduction reactions separately.

Electrodes in electrochemical setups are classified as anodes on which oxidation reactions occur, and as cathodes on which reduction reactions occur, both of which can be used as plasmonic photoelectrodes. The structures and working principles of photoanodes (metal or metal/n-type semiconductor) and photocathodes (metal or metal/p-type semiconductor) are briefly explained, as both are prevalent in the literature [106,107].

Heterostructures as a metal on an n-type semiconductor have been extensively investigated as photoanodes in plasmonic photocatalysis. For n-type semiconductors, Fermi levels are near the conduction band. When semiconductors come into contact with metals, they donate electrons to the metal to equalize Fermi levels. An upward band bending occurs at metal/semiconductor interface (called a "Schottky barrier"). When hot electrons created in the metal have sufficient energy, they can overcome the Schottky barrier to reach the conduction band of the semiconductor, as depicted in Fig. 18(a) [106]. However, due to upward band bending, they are rapidly swept from the interface, into the semiconductor. Such heterostructures are useful as a means of extending the lifetime of energetic electrons, which is longer in a semiconductor than in a metal. Electrons then travel along the external circuit, to accumulate at the counter electrode where they participate in reduction reactions, *e.g.*, water reduction, as shown in Fig. 18(a). Hot holes are left on metal surfaces to drive oxidation reactions, *e.g.*, water oxidation, also shown Fig. 18(a), resulting in an anodic photocurrent on the working electrode.

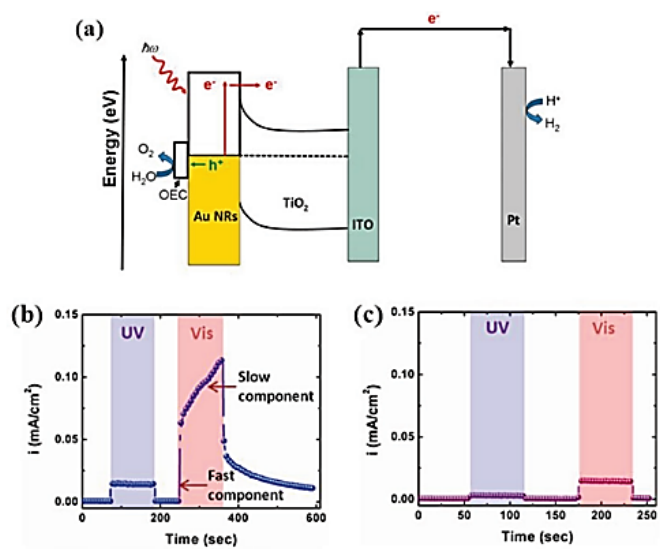

Fig. 18. Plasmonic metal/n-type semiconductor photoanode working electrode for redox of water. (a) Energy band diagram of a Au/TiO$_2$ photoanode. Absorption of visible light in Au generates energetic electron-hole pairs therein. Hot electrons emitted from Au into TiO$_2$ enable water reduction on the counter electrode (acting as cathode), and hot holes in Au are extracted by a Co-based oxygen evolution catalyst (Co-OEC) on Au to enable water oxidation. Current *vs.* time for Au/TiO$_2$ (b) with Co-OEC, and (c) without Co-OEC. Reprinted with permission from [106]. Copyright 2012 American Chemical Society.

Plasmonic photocathodes, consisting of metal/p-type semiconductor heterostructures, on which plasmon-generated hot holes are captured by p-type semiconductors, and hot electrons extracted from the metal drive reduction reactions, have also been reported, as sketched in Fig. 19(a) [107]. Theoretically, photoexcited holes above the interband edge in Au are considerably hotter than their photoexcited electron counterparts, which implies a greater collection efficiency for hot holes than hot electrons, for similar Schottky barrier heights. However, due to the relatively short mean free path of hot holes in a metal and the paucity of p-type semiconductors with wide bandgaps, harvesting hot holes from plasmonic metals remains challenging. p-GaN has recently been used as the semiconductor in plasmonic photocathodes, as sketched in Fig. 19(a). In Au/p-GaN heterostructures, excited holes in the Au nanoparticles transfer into the valence band of p-

GaN. During plasmon-driven $CO_2$ reduction, electrons trapped in the Au nanoparticles contribute to the cathodic photocurrent, as shown in Fig. 19(b).

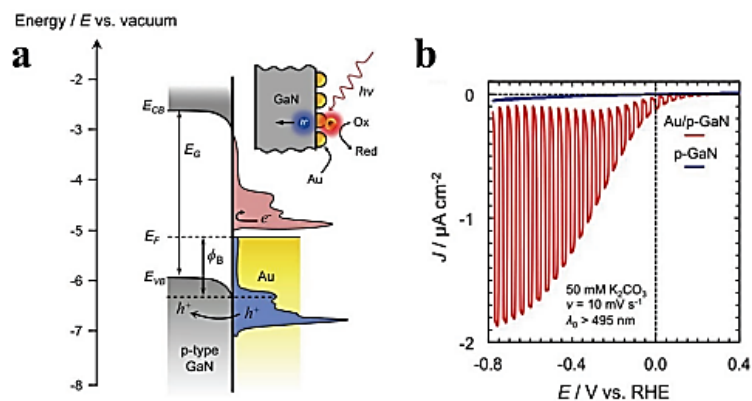

Fig. 19. Plasmonic metal/p-type semiconductor photocathode and reduction of $CO_2$. (a) Energy band diagram of Au/p-GaN photocathode, depicting the valence band edge $E_{VB}$, the conduction band edge $E_{CB}$, the bandgap energy $E_G$, the Fermi level energy $E_F$ and the Schottky barrier potential $\phi_B$. Plasmon excitation generates hot electrons (red) and hot holes (blue) above and below $E_F$, respectively. Only hot holes with energies larger than $e\phi_B$ can overcome the Schottky barrier and populate available valence band levels in p-GaN. (b) Linear sweep voltammetry of Au/p-GaN (red) and bare p-GaN (blue) photocathodes. Reprinted with permission from [107]. Copyright 2018 American Chemical Society.

Creating energetic carriers through SPP absorption on metals increases the local temperature, with heat diffusing into the nearby reaction volume. Due to the temperature dependence of electrochemical reactions, separating the roles of temperature and energetic carriers is not trivial yet essential to provide an understanding of results [108-113].

An analytical and experimental investigation of thermal effects on the electrochemical response of electrodes bearing plasmonic nanostructures under illumination was reported in [113]. In terms of time-dependent temperature profiles in the vicinity of illuminated electrodes, a straightforward approach was taken that considers heat flow via conduction away from a planar electrode solution interface, yielding analytical expressions for the temperature profile in the system and subsequent increases in electrochemical rate because of enhanced diffusion [113].

Fig. 20 illustrates the distribution of velocity and temperature in a system consisting of a cylindrical glassy carbon electrode surrounded by an insulating sheath, immersed in a cylindrical electrochemical cell containing an aqueous electrolyte, with a power input of 10 Wcm$^{-2}$ applied to the electrode over 10 s [113]. Local heat input at the electrode surface results in significant solution flow that runs vertically from the electrode edge, dissipating and changing direction further into the solution. Convective flows have the effect of lowering the temperature at the electrode surface by a factor of ~2 as they serve as a mechanism for transferring heat away from the electrode surface.

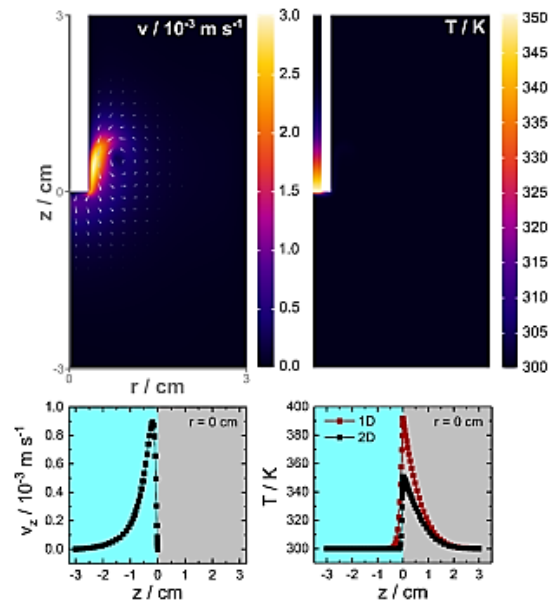

Fig. 20. Example velocity (left) and temperature (right) profiles near a 1.5 mm radius glassy carbon electrode surrounded by an insulating sheath (white rectangle) in an electrochemical cell. Finite element simulations of heat transfer including conduction and convection in water with $10\ W\,cm^{-2}$ heat input at the surface. Magnitude and temperature data are encoded using color maps, and values are given 10 s after the onset of heating. The bottom panels show the vertical profiles of the z-component of solution velocity, $v_z$, and of the temperature. The red curve on the bottom right panel

shows the temperature profile computed in 1D ignoring convection. Reprinted with permission from [113]. Copyright 2019 American Chemical Society.

Increasing local temperatures affect electrochemical reaction rates through enhancements in mass transfer due to convection and by altering the diffusion coefficients of the redox species. Furthermore, heterogeneous electron transfer rates and redox potentials are temperature dependent. As predicted by the theoretical analyses of systems including heat transfer by conduction and convection, it is expected that light absorption at electrode surfaces would produce local temperature increases and solution flows. Mass transfer enhancement alone would result in significant current increases, and these increases would apply to any electrochemical reaction involving dissolved reactants and/or products in the system concerned.

Models of diffusion, convection, and mass transport predict that redox currents increase approximately linearly with heating power rather than exponentially (as might be expected from the Arrhenius law), and that the current rises due to convection within 10 s of heating (*e.g.*, [113]). Thus, temperature trends are not straightforward, so electrochemical cells should be stabilized, electrode temperatures monitored, independent thermal control experiments conducted, and optical variables besides intensity should be varied to separate the effects of temperature from those of energetic carriers.

Most plasmonic catalysis research has involved colloidal arrangements of Au nanoparticles illuminated at visible wavelengths. This scenario, however, poses certain challenges: The temperature near nanoparticles is often difficult to predict and measure because of collective effects [110]. Furthermore, carriers excited by wavelengths above the interband threshold ($h\upsilon \sim$ 2 eV) have very short lifetimes (due to electron-electron scattering at high carrier energies) [114].

Alternatively, electrochemistry can be carried out with lithographically defined and evaporated Au microelectrodes on a substrate [115-117], provided the electrodes are annealed prior to use [115]. Such electrodes are shaped as a stripe and used simultaneously as a surface plasmon waveguide supporting Bloch LRSPPs at infrared wavelengths excited by grating couplers, as sketched in Fig. 21 [43, 116, 117]. The use of infrared wavelengths ensures that energetic carriers created in the stripe by LRSPP absorption are long lived. Counter electrodes can also be defined on chip, enabling an integrated plasmonic/electrochemical sensing chip.

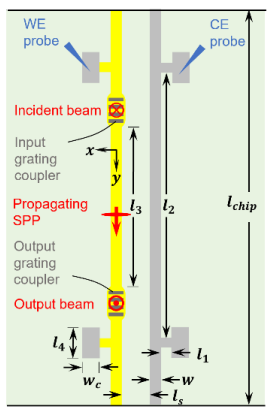

Fig. 21. Schematic in top view of the sensing chip. The thickness of the Au working electrode (WE, yellow) and of the Pt counter electrode (CE, gray) is $t = 35$ nm and their dimensions are: $l_1 = 29$ μm, $l_2 = 2600$ μm, $l_3 = 1850$ μm, $l_4 = 250$ μm, $l_{chip} = 3000$ μm, $w_c = 100$ μm, $w = 5$ μm, $l_s = 40$ μm. The Bloch LRSPPs propagating along the WE are excited using an optical beam normally incident on an input grating coupler. The output grating coupler produces a normally emerging output optical beam that can be monitored. Reprinted with permission from [117]. Copyright 2022 American Chemical Society.

The chip bearing a Au stripe waveguide also serving as a working electrode and a Pt counter electrode was integrated into a three-electrode electrochemical cell using an external reference electrode, and cyclic voltammograms were obtained while varying the incident optical power and wavelength, as shown in Fig. 22(a) [116]. By studying oxidation and reduction reactions separately, energetic hole processes were separated from energetic electron processes. Excitation of LRSPPs increased the redox current density by 10×, as shown in Fig. 22(b). The oxidation, reduction and equilibrium potentials dropped by as much as 2× and split in correlation with the photon energy beyond a clear threshold with SPP power, as shown in Fig. 22(c). Electrochemical impedance spectroscopy measurements showed that under LRSPP excitation, the charge transfer resistance was almost twice as low. During LRSPP excitation, the working electrode temperature was monitored *in situ* and in real-time, and independent control experiments eliminated thermal effects. An analysis of chronoamperometry results, with LRSPP on-off modulated at 600 Hz, shows a rapid current response modulated at the same frequency, which excludes thermally enhanced mass transfer. The opening of non-equilibrium redox channels associated with the transfer of energetic carriers to the redox species was invoked to explain these observations [116].

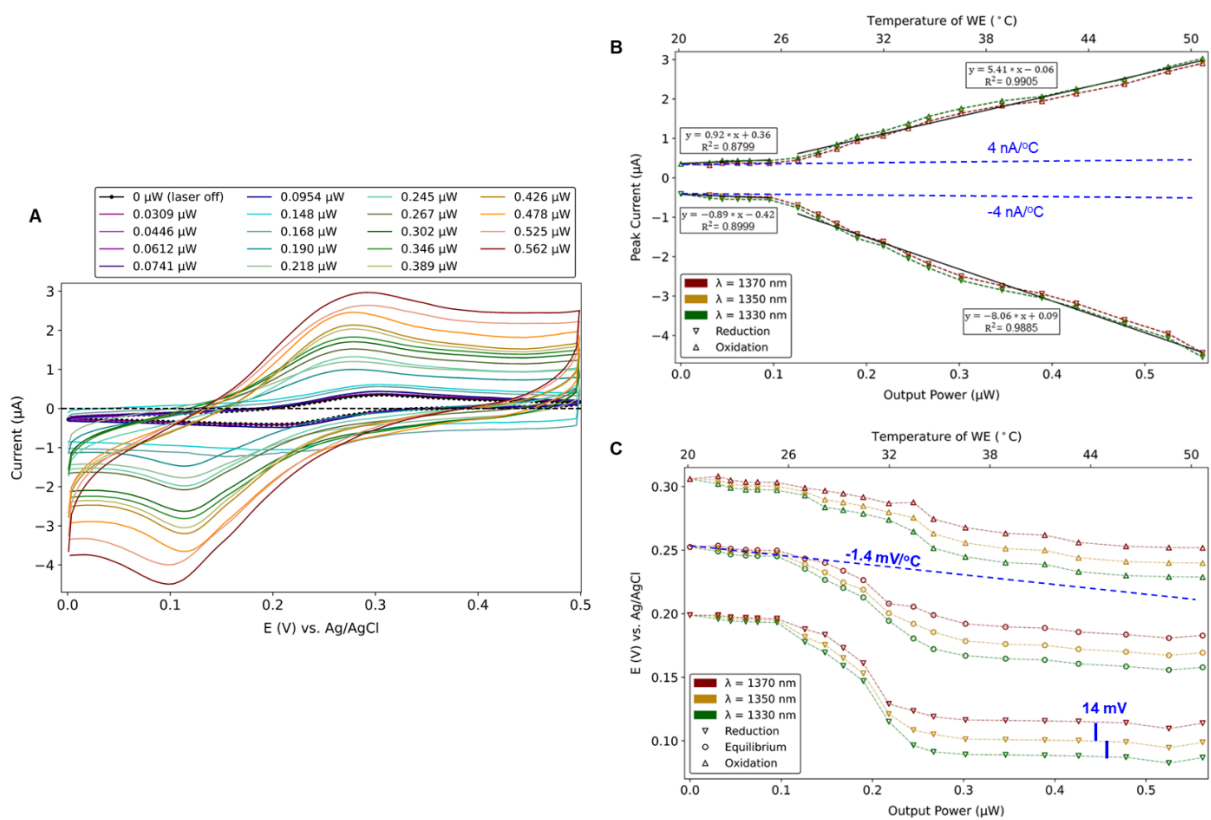

Fig. 22. Cyclic voltammetry under optical illumination. (a) CV curves obtained on a Au WE, in 0.5 mM $K_3[Fe(CN)_6]$ + 100 mM $KNO_3$ electrolyte, at a scan rate of 100 mV/s, for increasing output optical power (legend) at $\lambda_0$ = 1350 nm. The incident optical power ranged from 0 to 6.3

mW. The reference CV curve (laser off) is plotted as black dots. (b) Redox current peaks, and (c) potentials *vs*. output optical power, from CV curves measured at $\lambda_0$ = 1330, 1350 and 1370 nm. Linear thermal trends, measured independently, are added as the blue dashed lines to Parts (b) and (c), and rule out thermal effects. Reprinted with permission from Hirbodvash, Z.; Krupin, O.; Northfield, H.; Olivieri, A.; Baranova, E. A.; Berini, P. Infrared surface plasmons on a Au waveguide electrode open new redox channels associated with the transfer of energetic carriers. Sci. Adv., 2022, 8, eabm9303; DOI: 10.1126/sciadv.abm9303. Copyright 2022 The Authors, licensed under a Creative Commons Attribution (CC BY) license.

The output optical power from a waveguide electrode supporting Bloch LRSPPs was found to be proportional to the time-convolution of the electrochemical current density [117], enabling real-time convolutional electrochemistry. The optical response of a waveguide working electrode was derived theoretically, and validated experimentally using chronoamperometry and cyclic voltammetry measurements at various concentrations of potassium ferricyanide in potassium nitrate electrolyte [117]. By increasing the optical power, the LRSPP no longer acts solely as a probe of electrochemical activity, but also as a pump to create energetic electrons and holes, leading to significantly enhanced currents in this regime (*cf*. Fig. 22). The output optical power remains proportional to the time convolution of the current density in this high-power regime, even when energetic carriers are responsible for the redox reactions [117].

## 4. Conclusions

We introduced in a tutorial-like fashion basic concepts related to SPPs on planar structures and to electrochemistry, and we reviewed representative literature on their integration. Integration is motivated by the prospect of multimodal biosensors where the strengths of both techniques can be leveraged. Other motivating factors include the use of SPPs to probe electrochemical activity, leading naturally to real-time convolutional voltammetry by monitoring the output optical signal, or the use of SPPs as a "pump" affecting electrochemical reactions. Pumping occurs through SPP absorption in the working electrode, leading to the creation of energetic (hot) electrons and holes that can transfer more readily to the redox species. Enhanced carrier transfer can lead to enhanced electrochemical currents or may open new redox channels (pathways).

Integration challenges remain on several fronts. For instance, clearly differentiating thermal effects from the effects of energetic carriers in plasmonic hot carrier electrochemistry, and developing robust fabrication methods that integrate on-chip sealed microfluidic channels with optical (plasmonic) working and counter electrodes. Non-trivially, three types of interfaces must be simultaneously integrated on-chip: (*i*) sealed fluidic interfaces to couple microfluidic channels to external tubing, (*ii*) optical structures to couple incident light to plasmonic working electrodes and extract output light therefrom, and (*iii*) integrating isolated electrical contacts to connect to working and counter electrodes. These motivating factors and challenges, and the promise of new and exciting applications, are driving a vigorous global expansion of the field of surface plasmon electrochemistry.